\documentclass[
    aps,12pt,final,notitlepage,
    oneside,onecolumn,nofootinbib,noshowpacs,
    nolongbibliography,centertags
]{revtex4-2}

\usepackage[english]{babel}

\usepackage{amsmath}

\usepackage[]{graphicx}
\graphicspath{{figures/}}

\usepackage{natbib}

\usepackage[usenames,dvipsnames]{xcolor} % link colors tweak
\usepackage[
    colorlinks = true,
    urlcolor = MidnightBlue,
    linkcolor = Red,
    citecolor = NavyBlue
]{hyperref}

\newcommand{{\yeas}}{{YEASa}}
\newcommand{{\E}}{{E_0}}

\newcommand{{\ethr}}{\varepsilon_{\text{thr.}}} % threshold energy
\newcommand{{\lnA}}{\langle\ln{A}\rangle} % <lnA>
\newcommand{{\xmax}}{x_{\text{max}}}      % Xmax
\newcommand{{\xobs}}{x_{\text{obs}}}
\newcommand{{\dE}}{{\Delta E_S}}          % SD energy deposit
\newcommand{{\Xmp}}{x_{\text{max}}^p}
\newcommand{{\XmFe}}{x_{\text{max}}^{\text{Fe}}}
\newcommand{{\XmExp}}{x_{\text{max}}^{\text{exp}}}
\newcommand{{\rhom}}{\rho_{\mu}}          % muon density
\newcommand{{\rhos}}{\rho_s}              % surfase density

\newcommand{{\qgs}}{{\textsc{qgsj}et01}}
\newcommand{{\qgsii}}{{\textsc{qgsj}et-II.04}}
\newcommand{{\epos}}{{\textsc{epos}}}
\newcommand{{\eposlhc}}{{\textsc{epos-lhc}}}
\newcommand{{\sibyll}}{{\textsc{sibyll}-2.1}}
\newcommand{{\fluka}}{{\textsc{fluka}}}
\newcommand{{\corsika}}{{\textsc{corsika}}}

\newcommand{{\usec}}{{~$\mu$s}}    % usec
\newcommand{{\sqrm}}{{~m$^2$}}     % m^2
\newcommand{{\sqrkm}}{{~km$^2$}}   % km^2
\newcommand{\depth}{{~g/cm$^2$}}   % g/cm^2
\newcommand{\dens}{{~g/cm$^3$}}    % g/cm^3
\newcommand{\degr}{^{\circ}}     % degrees

\begin{document}

\title{Energy Spectrum of Ultrahigh-Energy Cosmic Rays according to Data from Ground-Based Scintillation Detectors of the Yakutsk EAS Array}

\author{A.\,V.~Glushkov}
\email{glushkov@ikfia.ysn.ru}

\author{M.\,I.~Pravdin}

\author{A.\,V.~Saburov}

\affiliation{
    Yu.~G.~Shafer Institute of Cosmophysical Research and Aeronomy SB RAS, \\
    31 Lenin Ave., 677027 Yakutsk, Russia
}

%\date{}

\begin{abstract}
    Results obtained from an analysis of the energy spectrum of cosmic rays with energies in the region of $\E \ge 10^{17}$~eV over the period of continuous observations from 1974 to 2017 are presented. A refined expression for estimating the primary-particle energy is used for individual events. This expression is derived from calculations aimed at determining the responses of the ground-based and underground scintillation detectors of the Yakutsk array for studying extensive air showers (EAS) and performed within the \qgs, \qgsii, \sibyll{} and \eposlhc{} models by employing the \corsika{} code package. The new estimate of $\E$ is substantially lower than its counterpart used earlier.
\end{abstract}

\maketitle

\section{Introduction}

The energy spectrum of ultrahigh-energy cosmic rays (CR) ($\E \ge 10^{17}$~eV) is one of the key links in the chain of problems on the path toward obtaining deeper insight into the nature of primary particles that have such energies. Experimental results obtained at different arrays for studying extensive air showers (EAS)~\cite{bib:1, bib:2, bib:3, bib:4, bib:5, bib:6, bib:7} differ in absolute intensity nearly by a factor of two but are close in shape~\cite{bib:8}. This situation is due largely to the fact that, at the majority of large arrays worldwide, use is made of different methods for determining the primary-particle energy $\E$ in view of the difference of the procedures for EAS detection at these arrays. Here, one cannot dispense with invoking theoretical ideas of the development of EAS.

The Yakutsk EAS array is the oldest in the world. It has operated continuously since 1974,
standing out among the other large arrays owing to its multifunctionality-specifically, the ability to measure simultaneously all EAS particles with groundbased scintillation detectors of area 2~\sqrm, muons at a threshold above $1.0 \cdot \sec\theta$~GeV with analogous underground detectors, and Cherenkov light from EAS. The Cherenkov component carries information about approximately 80\% of the primary energy scattered by a shower in the Earth's atmosphere and makes it possible to determine $\E$ calorimetrically~\cite{bib:9, bib:10, bib:11, bib:12}. For the first time ever, this method was applied in~\cite{bib:13} at energies around $10^{15}$~eV. At the Yakutsk EAS array, it was implemented in the energy range of $\E \simeq (1.0-100) \times 10^{17}$~eV and the zenith-angle range of $\theta \le 45\degr$~\cite{bib:9}:

\begin{gather}
    \E = (4.1 \pm 1.4) \times 10^{17}(S_{600}(0\degr))^{0.97 \pm 0.04}~\text{[eV],}
    \label{eq:1} \\
    S_{600}(0\degr) = S_{600}(\theta)
        \exp\left(
            \frac{(\sec\theta - 1)\cdot 1020}{\lambda}
        \right)~\text{[m}^{-2}\text{],}
        \label{eq:2} \\
    \lambda = 400 \pm 45~\text{[g/cm}^2\text{].}
    \label{eq:3}
\end{gather}

\noindent
Here, $S_{600}(\theta)$ is the particle density measured by ground-based scintillation detectors at the distance of $r = 600$~m from the shower axis. Later, relations (\ref{eq:1}) and (\ref{eq:3}) changed somewhat to become~\cite{bib:10, bib:11, bib:12}:

\begin{gather}
    \E = (4.8 \pm 1.6) \times 10^{17} (S_{600}(0\degr))^{1.00 \pm 0.02}~\text{[eV],}
    \label{eq:4} \\
    \lambda = (450 \pm 44) + (32 \pm 15) \lg(S_{600}(0\degr))~\text{[g/cm}^2\text{].}
    \label{eq:5}
\end{gather}

The cosmic-ray energy spectrum estimated on the basis of expression (\ref{eq:4}) proved to be substantially higher in intensity than all data obtained worldwide. In~\cite{bib:14, bib:15}, we revisited the energy calibration of showers by means of the \corsika{} code~\cite{bib:16} on the basis of modern hadron-interaction models considered below.

\section{Evaluating primary energy}

\subsection{Data on Lateral Distribution from Scintillation Detectors}

Basic parameters of EAS at the Yakutsk EAS array (such as arrival direction, coordinates of the axis, and primary energy) are determined with the aid of the lateral distribution of all particles (electrons, muons, and high-energy photons) recorded by ground-based scintillation detectors. These particles traverse a multilayered shield from snow, iron, wood, and duralumin (the total thickness is about 2.5~\depth) and thereupon a scintillator 5~cm thick (its density is 1.06~\dens), where they deposit some energy $\dE(r)$, which is proportional to the number of particles that traversed the detector. In practice, this energy deposition is measured in relative units; that is,

\begin{equation}
    \rho_S(r) = \frac{\dE(r)}{E_1}~\text{[m}^{-2}\text{],}
    \label{eq:6}
\end{equation}

\noindent
where $E_1 = 11.75$~MeV is the energy deposited in a ground-based detector upon the passage through it of one vertical relativistic muon (unit response).

The scintillation detectors are calibrated and are controlled with the aid of the amplitude density spectra from background cosmic-ray particles~\cite{bib:17}. In doing this, use is made of integrated spectra of two types. Of them, the first is the spectrum from one of
the detectors controlled by the neighboring detector from the same station (spectrum of ``double coincidences'' with a frequency of about 2 to 3~s$^{-1}$). The second is a spectrum without a control; the respective frequency is about 200~s$^{-1}$. It is used to calibrate muon detectors. Both spectra have a power-law form; that is,

\begin{equation}
    F(>\rho) \sim \rho^{-\eta} \sim U^{-\eta}\text{,}
    \label{eq:7}
\end{equation}

\noindent
where $\eta = 1.7$ and $3.1$ in, respectively, the first and the second case and $\rho = U/U_1$ is the particle density in units of the amplitude $U_1$ of the signal of the reference
detector from vertical relativistic cosmic muons. The procedure of calibration and control reduces to monitoring the quantity $U_1$ for all detectors by periodically measuring their density spectra. This is done once per two days, the double-coincidence spectra and spectra without control being taken for two hours and 30~minutes, respectively.

We have calculated lateral distributions of responses on the basis of the \qgs~\cite{bib:18}, \qgsii~\cite{bib:19}, \sibyll~\cite{bib:20}, and \eposlhc~\cite{bib:21} models for primary protons and iron nuclei in the range of energies between $10^{17.0}$ and $10^{19.5}$~eV for various zenith angles. As a model for low energies, we took \fluka~\cite{bib:22}. First, we calculated the responses $u_k(E, \theta)$ to single particles of type $k$ (here, $k$ is an electron, a muon, or a photon) with energy $E$. In doing this, we took into account all processes of energy deposition and absorption in the shield and in the scintillator and the cross sections for the interactions undergone by these particles. After that, the development of EAS in the real atmosphere was simulated with the aid of the \corsika{} code. Five hundred showers were generated for each set of primary parameters (including primary-particle mass, primary energy, and zenith angle). With the aim of reducing the computer time, we invoked the thin-sampling mechanism, its parameters being $E_i / \E = 10^{-5}$ and $w_{\text{max}} = 10^4$. A rescale to the density was accomplished upon taking into account the number of particles per detector of given area. Averaging over respective showers was performed, and the energy spectra $d_k(E, r, \theta)$ were calculated for all types of particles in the intervals $(\lg{r_j}, \lg(r_j + 0.04))$ of distances. The signal in (\ref{eq:6}) was determined by the sum of the responses; that is,

\begin{equation}
    \rho_S(r) = \sum_{k - 1}^{3} \sum_{i = 1}^{I_k}
    u_k(E_i , \theta_i) \cdot d_k(E_i, r, \theta_i)\text{,}
    \label{eq:8}
\end{equation}

\noindent
where $I_k$ is the number of particles that belong to type $k$ and which hit the detector.

\begin{figure}
    \centering
    \includegraphics[width=0.65\textwidth]{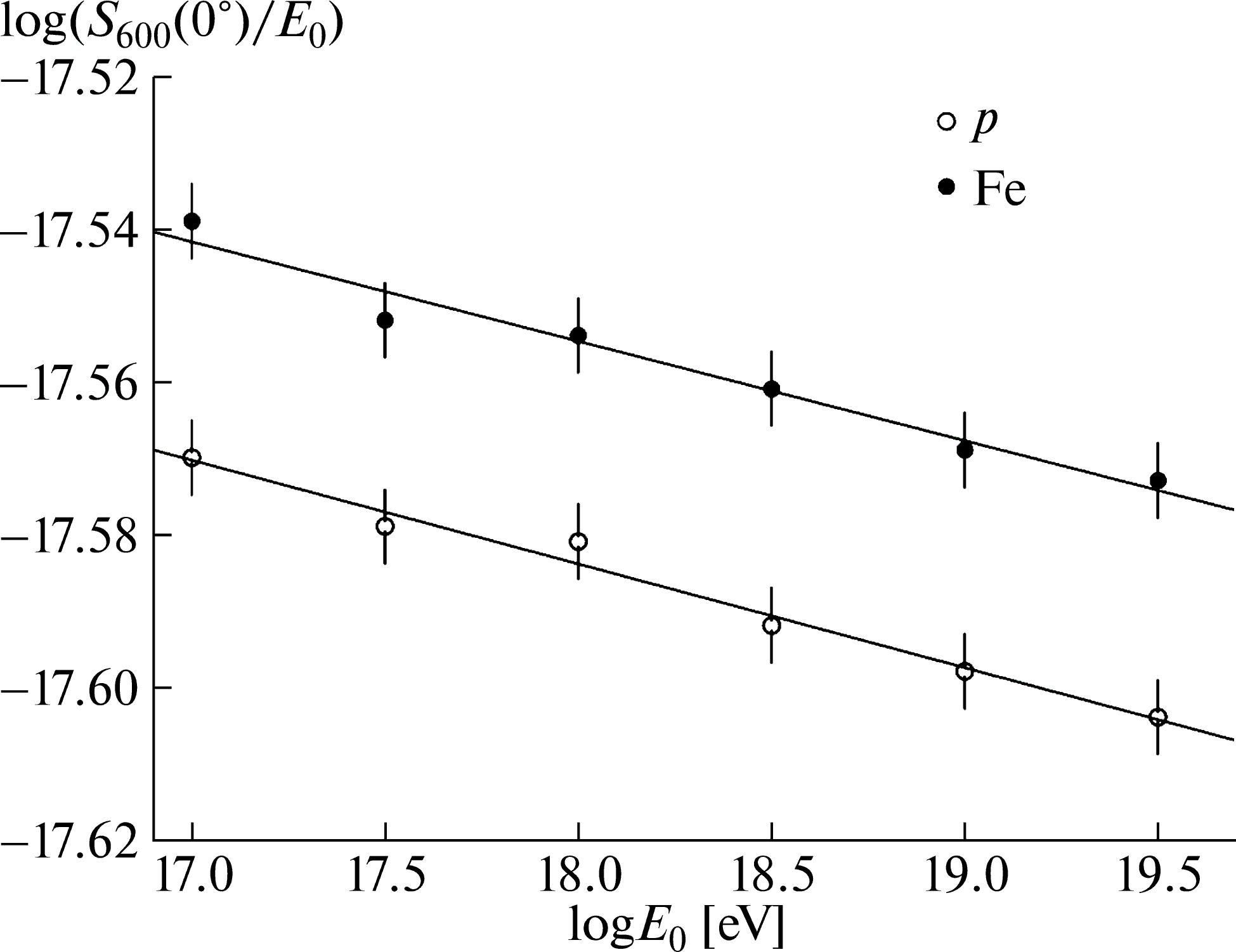}
    \caption{$\lg(S_{600}(0\degr)/\E)$ as a function of the energy $\E$ for primary (open circles) protons and (closed circles ) iron nuclei in vertical showers according to the \qgs{} model. The curves represent linear approximations of the experimental points.}
    \label{fig:1}
\end{figure}

Figure~\ref{fig:1} shows $\lg(S_{600}(0\degr)/\E)$ as a function of $\E$ for primary (open circles) protons and (closed circles) iron nuclei according to calculations on the
basis of the \qgs{} model. These values satisfy the relation

\begin{equation}
    \E = (3.55 \pm 0.1) \times 10^{17} \cdot (S_{600}(0\degr))^{1.02}~\text{[eV].}
    \label{eq:9}
\end{equation}

\noindent
The estimates based on the application of the \qgsii, \eposlhc, and \sibyll{} models are, respectively,

\begin{gather}
    \E = (3.19 \pm 0.1) \times 10^{17} \cdot (S_{600}(0\degr))^{1.03}~\text{[eV],}
    \label{eq:10} \\
    \E = (2.87 \pm 0.1) \times 10^{17} \cdot (S_{600}(0\degr))^{1.03}~\text{[eV],}
    \label{eq:11} \\
    \E = (3.72 \pm 0.1) \times 10^{17} \cdot (S_{600}(0\degr))^{1.02}~\text{[eV].}
    \label{eq:12}
\end{gather}

\begin{figure}
    \centering
    \includegraphics[width=0.65\textwidth]{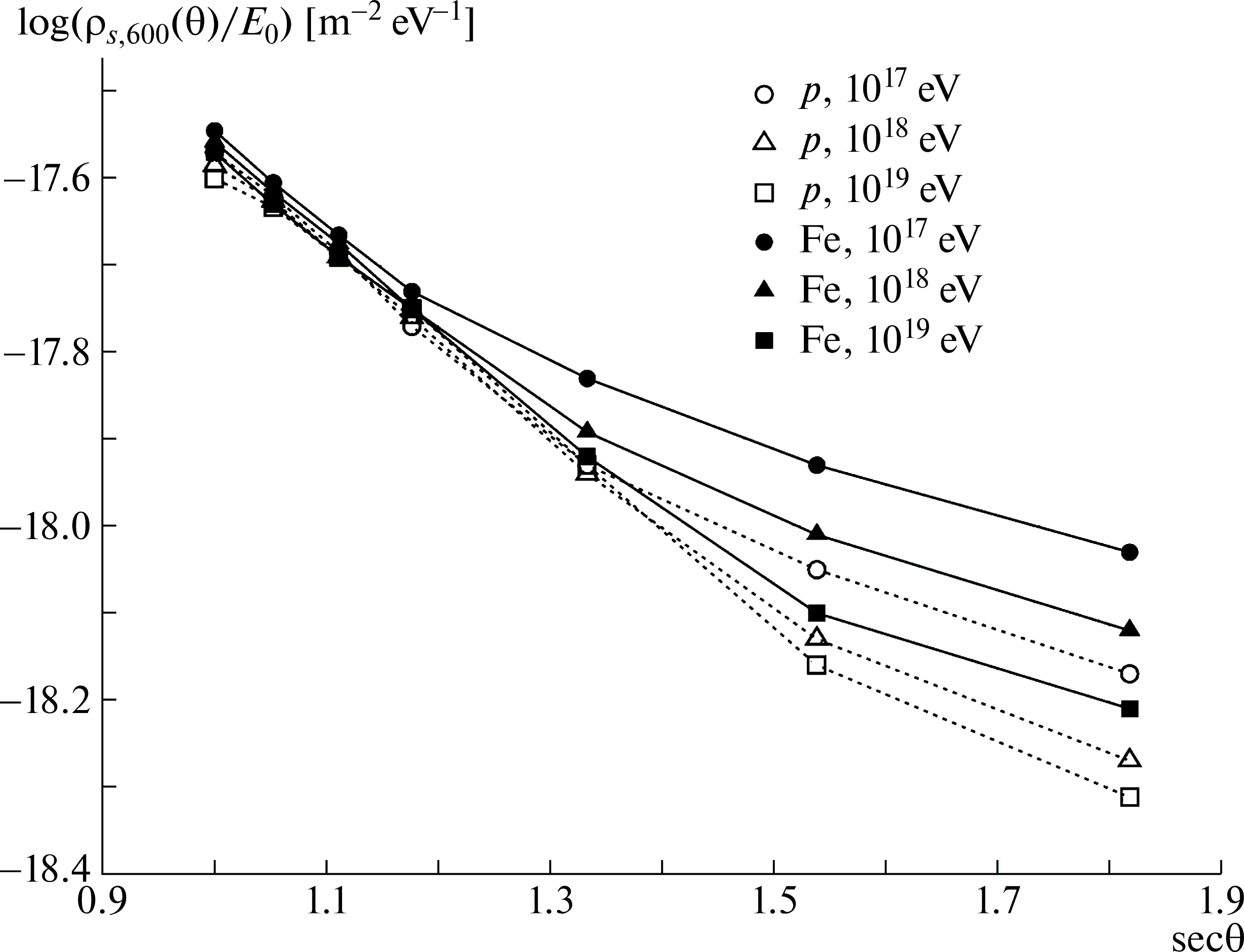}
    \caption{$\lg(S_{600}(\theta)/\E)$ as a function of the zenith angle according to calculations performed on the basis of the \qgs{} model for primary (open symbols) protons and (closed symbols) iron nuclei of energy $E =$ (open and closed circles) $10^{17}$ (open and closed triangles) $10^{18}$ and (open and closed boxes) $10^{19}$~eV. The points on display were connected by lines in order to guide the eye.}
    \label{fig:2}
\end{figure}

\begin{figure}
    \centering
    \includegraphics[width=0.65\textwidth]{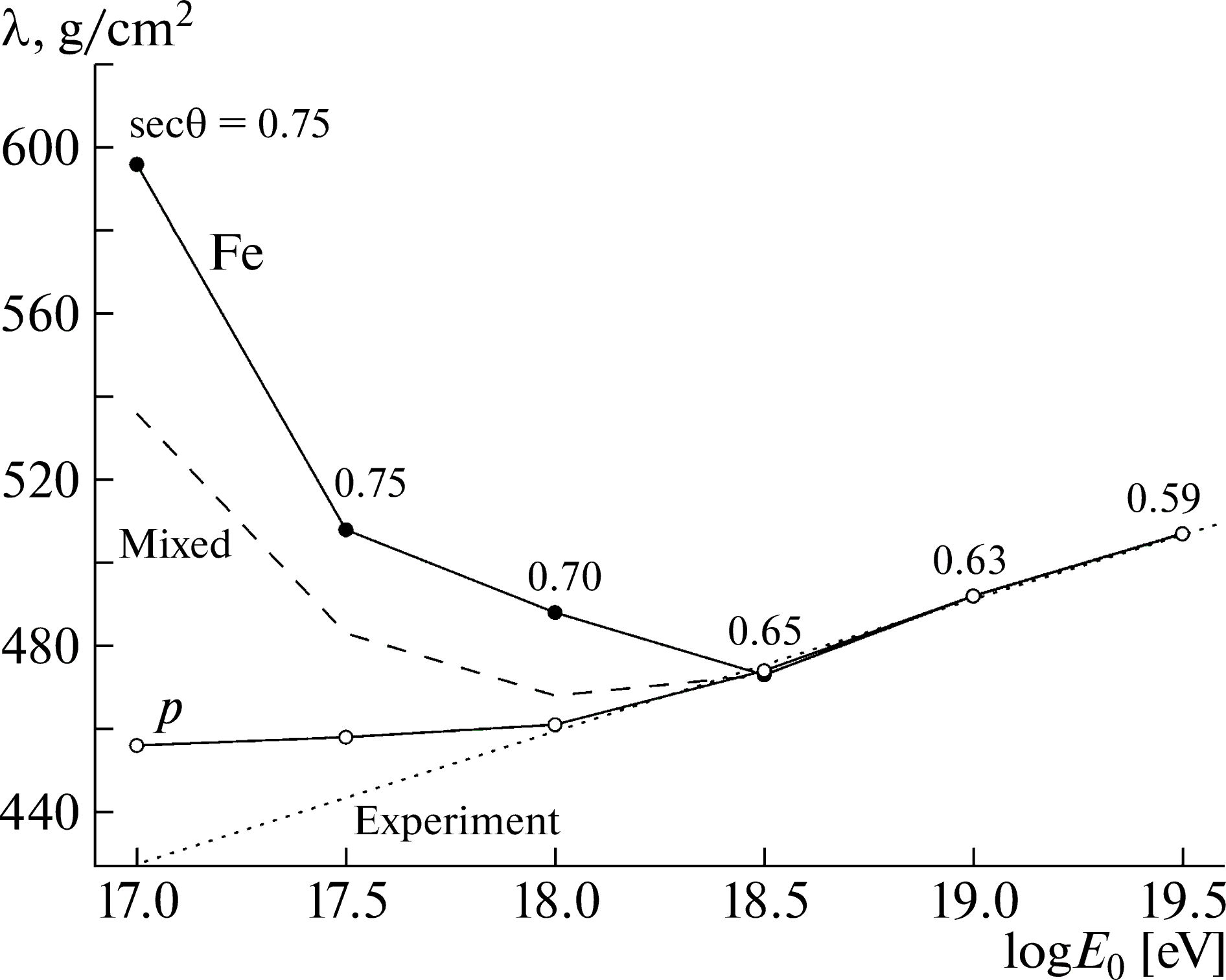}
    \caption{Absorption ranges in (\ref{eq:2}) upon rescaling $S_{600}(\theta)$ from inclined to vertical showers according to the \qgs{} model for primary protons ($p$), mixed
    composition, and iron nuclei (Fe) versus $\E$. The numbers indicate the limiting admissible zenith angles. The dotted curve represents relation (\ref{eq:5}).}
    \label{fig:3}
\end{figure}

Figure~\ref{fig:2} gives $\lg(S_{600}(\theta)/\E)$ as a function of the zenith angle according to calculations on the basis of the \qgs{} model. This dependence corresponds to the variations in $\lambda$ in (\ref{eq:2}) that are shown in Fig.~\ref{fig:3}. The dashed curve in Fig.~\ref{fig:3} represents absorption ranges for a mixed composition of primary
nuclei according to our experimental data from~\cite{bib:23, bib:24}. The dotted curve corresponds to the empirical relation (\ref{eq:5}).

\subsection{Data Obtained Calorimetrically}

The method in question is described here by considering the example of experimental data from~\cite{bib:9, bib:10} taken as the basis in developing a calorimetric method for estimating $\E$ at the Yakutsk EAS array. Tables~\ref{t:1} and \ref{t:2} give observed parameters and basic constituents of $\E = 10^{18}$~eV in showers characterized by $\cos\theta = 0.95$. The ``average $p$--Fe'' line corresponds to values averaged over the CR composition and over all models. The electron–photon component energy scattered in the atmosphere is

\begin{equation}
    E_i = E_{\gamma} + E_{\text{ion.}}\text{,}
    \label{eq:13}
\end{equation}

\noindent
where $E_{\gamma}$ is the gamma-ray energy at the observation level and $E_{\text{ion.}}$ is the total ionization loss of all electrons. This loss is proportional to the total flux
of Cherenkov light, $F$, in the atmosphere; that is,

\begin{equation}
    E_i = k \cdot F\text{,}
    \label{eq:14}
\end{equation}

\noindent where

\begin{equation}
    k = k_{\gamma} + k_{\text{ion.}} =
    \frac{E_{\gamma} + E_{\text{ion.}}}{F}~\text{[eV/photon eV}^{-1}\text{].}
    \label{eq:15}
\end{equation}

\begin{table}
    \caption{Observed parameters of EAS characterized by $E = 10^{18}$~eV and $\cos\theta = 0.95$ and initiated by primary nuclei ($A$) according to the \corsika{} code~\cite{bib:16} and according to the experiments reported in~\cite{bib:9, bib:10}}.
    \label{t:1}
    \renewcommand{\arraystretch}{0.75}
    \begin{tabular}{llrrrrrr}
        \hline
        \hline
        Model & $A$ & $k_{\gamma}(\theta)$, eV$^2$ &
        $k_{\text{ion.}}(\theta)$, eV$^2$ & $F(\theta)$, eV$^{-1}$ &
        $N_S(\theta)$ & $S_{600}(\theta)$, m$^{-2}$ & $N_{\mu}(\theta)$ \\
              &     & $(\times 10^4)$ & $(\times 10^4)$ & $(\times 10^{13})$ &
                      $(\times 10^8)$ &    & $(\times 10^6)$ \\
        \hline
        \qgs & $p$ & 0.341 & 2.846 & 2.104 & 2.178 & 2.312 & 5.000 \\
             & Fe  & 0.224 & 2.910 & 2.148 & 1.250 & 2.432 & 7.225 \\
        \qgsii & $p$ & 0.364 & 2.816 & 2.070 & 2.296 & 2.438 & 5.582 \\
               & Fe  & 0.246 & 2.894 & 2.148 & 1.358 & 2.636 & 7.777 \\
        \sibyll & $p$ & 0.345 & 2.822 & 2.100 & 2.512 & 2.193 & 4.254 \\
                & Fe  & 0.224 & 2.910 & 2.228 & 1.384 & 2.249 & 4.930 \\
        \eposlhc & $p$ & 0.377 & 2.815 & 2.023 & 2.355 & 2.655 & 5.905 \\
                 & Fe  & 0.230 & 2.894 & 2.133 & 1.419 & 2.917 & 8.180 \\
        \hline
        Average & $p$     & 0.357 & 2.825 & 2.074 & 2.335 & 2.400 & 5.185 \\
        Average & Fe      & 0.231 & 2.902 & 2.164 & 1.353 & 2.558 & 7.028 \\
        Average & $p$--Fe & 0.294 & 2.864 & 2.119 & 1.844 & 2.479 & 6.107 \\
        \hline
        Experiment~\cite{bib:9, bib:10} &
        -- & \multicolumn{2}{c}{3.700} &
           2.510 & 1.793 & 2.656 & 6.000 \\
        \hline
        \hline
    \end{tabular}
\end{table}

\begin{table}
    \caption{Energy balance in EAS characterized by $\E = 10^{18}$~eV and $\cos\theta = 0.95$ and initiated by primary nuclei ($A$) according to the \corsika code~\cite{bib:16} and according to the experiments reported in~\cite{bib:9, bib:10}.}
    \label{t:2}
    \centering
    \renewcommand{\arraystretch}{0.75}
    \begin{tabular}{llrrrrrr}
        \hline
        \hline
        Model & $A$ & $E_{\gamma}$, eV & $E_{\text{ion.}}$, eV &
        $E_{\text{el.}}$, eV & $E_{\mu}$, eV & $\Delta E$, eV & $\E$, eV \\
              &     & $(\times 10^{17})$ & $(\times 10^{17})$ &
                      $(\times 10^{17})$ & $(\times 10^{17})$ &
                      $(\times 10^{17})$ & $(\times 10^{17})$ \\
        \hline
        \qgs & $p$ & 0.806 & 6.620 & 1.469 & 0.517 & 0.565 & 9.978 \\
             & Fe  & 0.529 & 6.600 & 1.306 & 0.785 & 0.798 & 9.972 \\
        \qgsii & $p$ & 0.859 & 6.476 & 1.474 & 0.547 & 0.624 & 9.980 \\
               & Fe  & 0.582 & 6.430 & 1.302 & 0.844 & 0.866 & 9.981 \\
        \sibyll & $p$ & 0.909 & 6.625 & 1.523 & 0.428 & 0.491 & 9.976 \\
                & Fe  & 0.528 & 6.679 & 1.340 & 0.702 & 0.716 & 9.965 \\
        \eposlhc & $p$ & 0.891 & 6.412 & 1.482 & 0.524 & 0.657 & 9.966 \\
                 & Fe  & 0.543 & 6.415 & 1.305 & 0.794 & 0.898 & 9.955 \\
        \hline
        Average & $p$     & 0.866 & 6.533 & 1.487 & 0.504 & 0.584 & 9.974 \\
        Average & Fe      & 0.546 & 6.531 & 1.313 & 0.781 & 0.820 & 9.968 \\
        Average & $p$--Fe & 0.706 & 6.532 & 1.400 & 0.643 & 0.702 & 9.970 \\
        \hline
        Experiment~\cite{bib:9, bib:10} & -- & \multicolumn{2}{c}{9.287} &
            0.947 & 0.636 & 0.860 & 11.730 \\
        New estimate & -- & \multicolumn{2}{c}{7.926} &
            0.947 & 0.618 & 0.702 & 10.190 \\
        \hline
        \hline
    \end{tabular}
\end{table}

\begin{figure}
    \centering
    \includegraphics[width=0.65\textwidth]{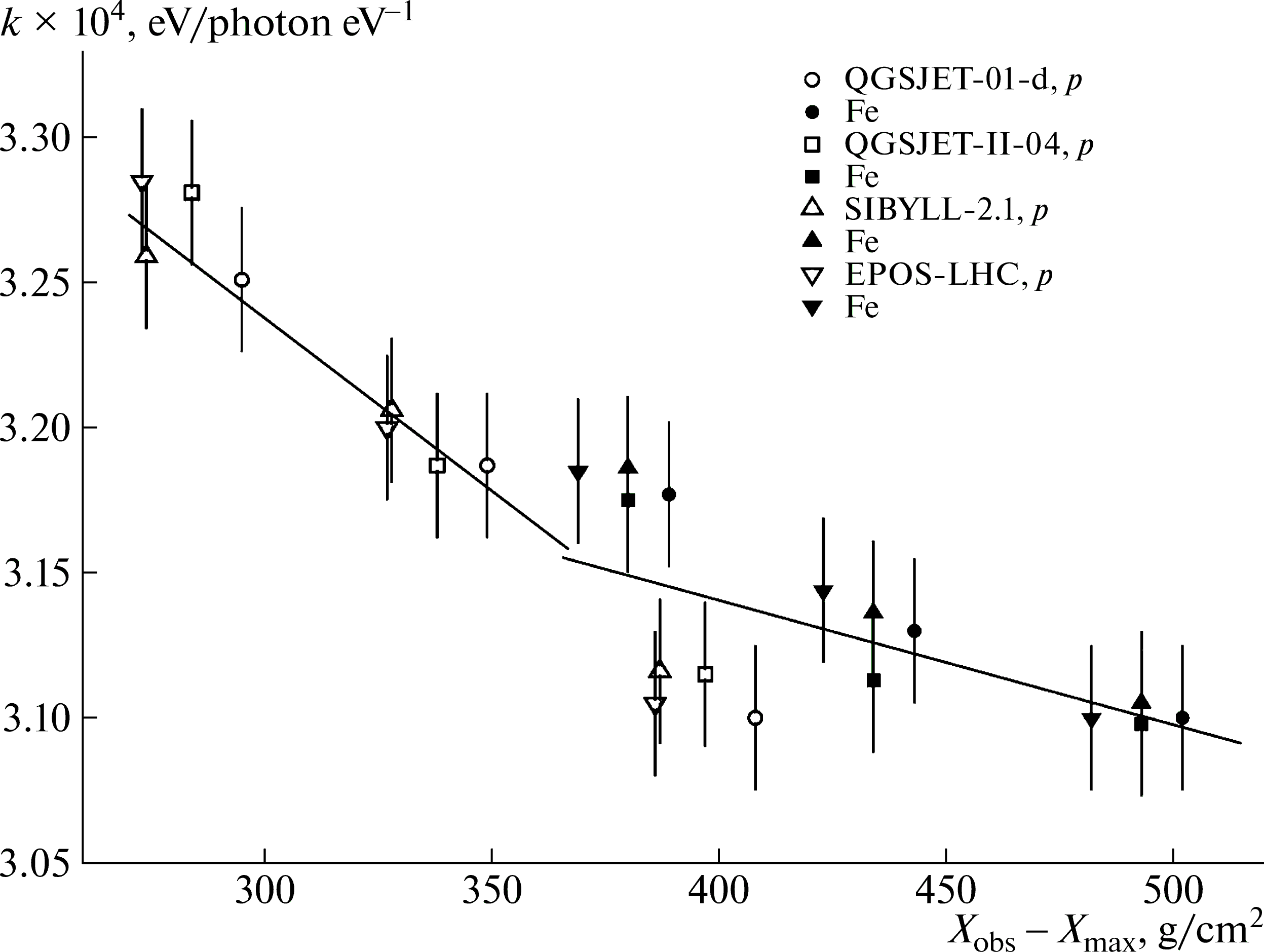}
    \caption{Rescaling coefficient in (\ref{eq:15}) as a function of the distance between the shower-maximum position $\xmax$ and the observation level $\xobs = 1020 \cdot \sec\theta$~\depth{} for primary (open symbols) protons and (closed symbols) iron nuclei according to the (open and closed circles) \qgs, (open and closed boxes) \qgsii, (open and closed triangles) \sibyll, and (inverted open and closed triangles) \eposlhc{} models. The lines represent approximations.}
    \label{fig:4}
\end{figure}

Figure~\ref{fig:4} shows the rescaling coefficient in (\ref{eq:15}) as a function of the distance between the shower maximum position $\xmax$ and the observation level $\xobs = 1020 \cdot \sec\theta$~\depth. The flux $F$ was found with allowance for its weakening by a factor of 1.15 because of Rayleigh light scattering in the absolutely clean atmosphere and a deterioration of its transparency by a factor of 1.1 for the shower sample from~\cite{bib:9, bib:10}. It is given within a 1-eV radiation interval; that is,

\begin{equation}
    F = 1.265 \cdot \frac{F_{\text{obs}}}{\Delta\epsilon}\text{,}
    \label{eq:16}
\end{equation}

\noindent
where $F_{\text{obs}}$ is the flux measured under conditions of the real experiment and

\begin{equation}
    \Delta\epsilon = 12400 \cdot
    \left(\frac{1}{\lambda_1} - \frac{1}{\lambda_2}\right) \simeq 2.58~\text{[eV].}
    \label{eq:17}
\end{equation}

\noindent
In the case being considered, we have $\lambda_1 = 3000$~\AA{} and $\lambda_2 = 8000$~\AA. The energy $E_{\text{el.}}$  is carried by the electron–photon component beyond the array plane. It was calculated by integrating the differential energy loss along the cascade curve $N_e(x)$ down to the observation level $\xobs$; that is,

\begin{equation}
    E_{\text{el.}} = \int_{\xobs}^{\infty}
        \frac{\text{d}E_i}{\text{d}x} N_e(x)\text{d}x \simeq
        2.2 \times 10^6 \cdot N_e(\xobs) \int_{\xobs}^{\infty}
        \exp{\frac{\xobs - x}{\lambda_a}}\text{d}x\text{,}
    \label{eq:18}
\end{equation}

\noindent
where $N_e(\xobs)$ is the number of electrons at the observation level. It was found from the relation

\begin{equation}
    N_e(\xobs) \simeq \left<N_S(\xobs)\right> - 1.8\left<N_{\mu}(\xobs)\right>
    \text{,}
    \label{eq:19}
\end{equation}

\noindent
where $\left<N_S(\xobs )\right>$ and $\left<N_{\mu}(\xobs)\right>$ are the average values of the total numbers of responses to, respectively, all particles and muons at a threshold above 1~GeV.

The muon energy $E_{\mu}$ was measured experimentally as

\begin{equation}
    E_{\text{mu}} \simeq \left<E_{1\mu}\right> \left<N_{\mu}(\xobs)\right>\text{,}
    \label{eq:20}
\end{equation}

\noindent
where $\left<E_{1\mu}\right> = 10.6$~GeV is the average energy of one muon.

From the calculated values in Table~\ref{t:2} that were averaged over all models, it follows that the total value $E_i + E_{el} + E_{\mu}$ is about 93\% of the primary energy. Its remaining part, $\Delta E$, is not controlled at the Yakutsk EAS array. It includes the neutrino energy, energy transferred to nuclei in various reactions and the muon and hadron energy losses by atmosphere ionization. In~\cite{bib:9, bib:10}, its value was taken from earlier calculations. Roughly, it is compatible with the estimates obtained
with the aid of the \corsika{} code~\cite{bib:16}.

\begin{figure}
    \centering
    \includegraphics[width=0.65\textwidth]{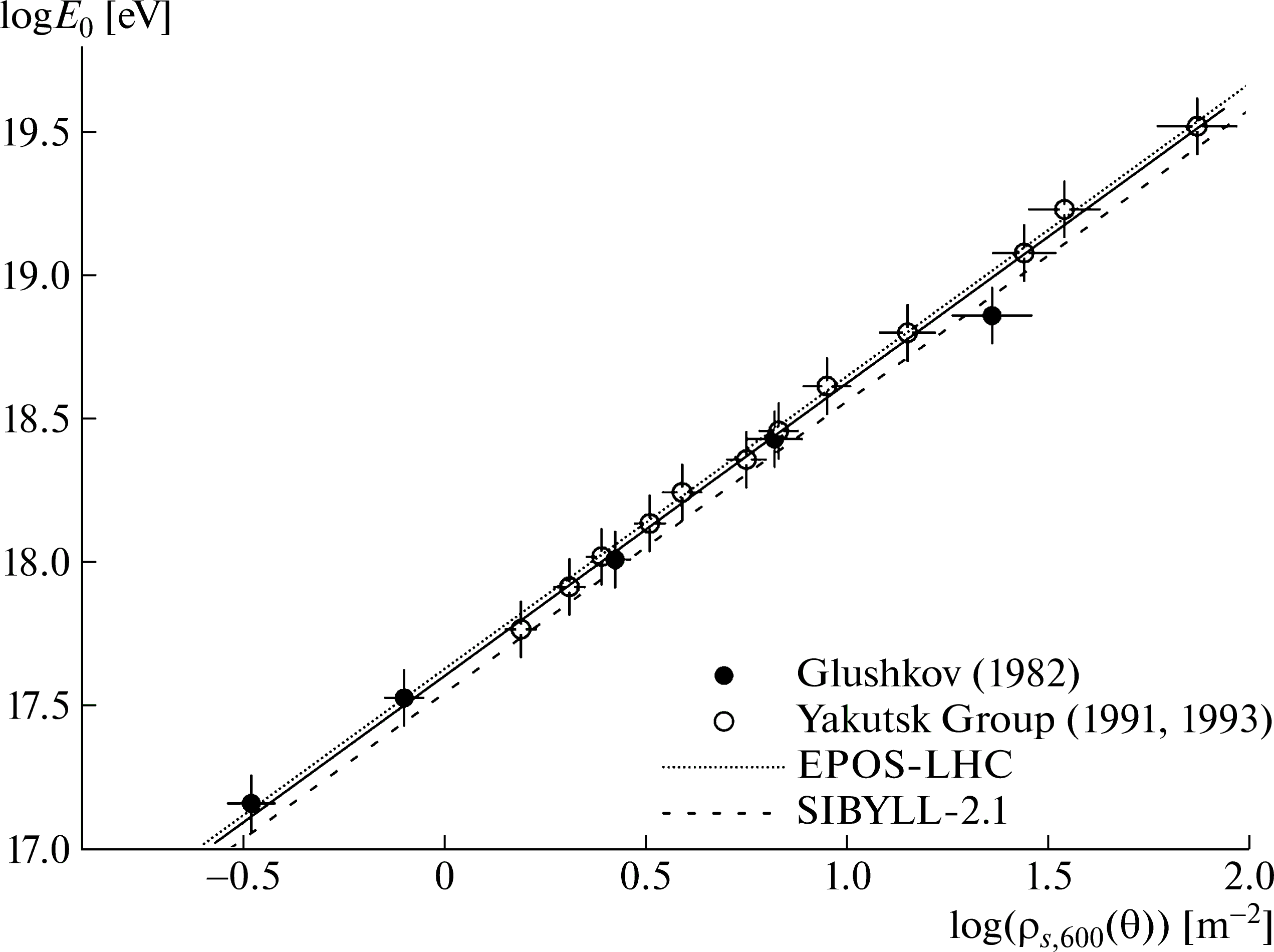}
    \caption{Primary energy $\E$ as a function of the parameter $S_{600}(\theta)$ in showers characterized by $\left<\cos\theta\right> = 0.95$ according to data from (closed circles)~\cite{bib:9} and (open circles)~\cite{bib:10} upon the application of the new calorimetric procedure (see main body of the text). The solid line stands for the best approximation of all data, while the dashed and dotted lines represent relations (\ref{eq:11}) and (\ref{eq:12}), respectively, for the above zenith angle.}
    \label{fig:5}
\end{figure}

The rightmost column of Table~\ref{t:2} contains the total values of all preceding components. The energy of $\E = 1.173 \times 10^{18}$~eV in the ``Experiment'' line exceeds its averaged model estimate $\left<\E\right> = 0.997 \times 10^{18}$~eV by a factor of about 1.177. This difference arose because of the use in~\cite{bib:9, bib:10} of the coefficient $k = 3.7 \times 10^4$~eV/photon~eV$^{-1}$ overestimated in relation to its calculated value of $\left<k\right> = 3.158 \times 10^4$ eV/photon~eV$^{-1}$. The new estimate $\E = 1.019 \times 10^{18}$~eV obtained by means of the above calorimetric method with the refined values of $E_i = \left<k\right>F$, $\left<E_{1\mu}\right> = 10.3$~GeV, and $\Delta E$ in the lowermost line of Table~\ref{t:2}. It is shown, along with other data from~\cite{bib:9}, in Fig.~\ref{fig:5} (closed circles). The open circles in this figure represent data from~\cite{bib:10} for which the values of $F$ and $E_{\text{ion.}}$ were modified via refining the transparency of the atmosphere and via employing the the new coefficient $k$ (see Fig.~\ref{fig:4}). The solid line corresponds to the dependence

\begin{equation}
    \E = (3.76 \pm 0.30) \times 10^{17} \cdot
    (S_{600}(0\degr))^{1.02 \pm 0.02}~\text{[eV],}
    \label{eq:21}
\end{equation}

\noindent
which complies with all experimental points upon rescaling $S_{600}(18.2\degr)$ to a vertical direction according to Eq.~(\ref{eq:2}) by employing the absorption length $\lambda$ represented by the dashed curve in Fig.~\ref{fig:3} (for a mixed composition of primary particles). The dashed and dotted lines in Fig.~\ref{fig:5} correspond to relations (\ref{eq:11}) and (\ref{eq:12}), which characterize the applicability limits for the models of EAS development that were considered above. The \qgs{} and \sibyll models provide the best agreement with experimental data.

\section{Primary energy spectrum}

\begin{figure}
    \centering
    \includegraphics[width=0.75\textwidth]{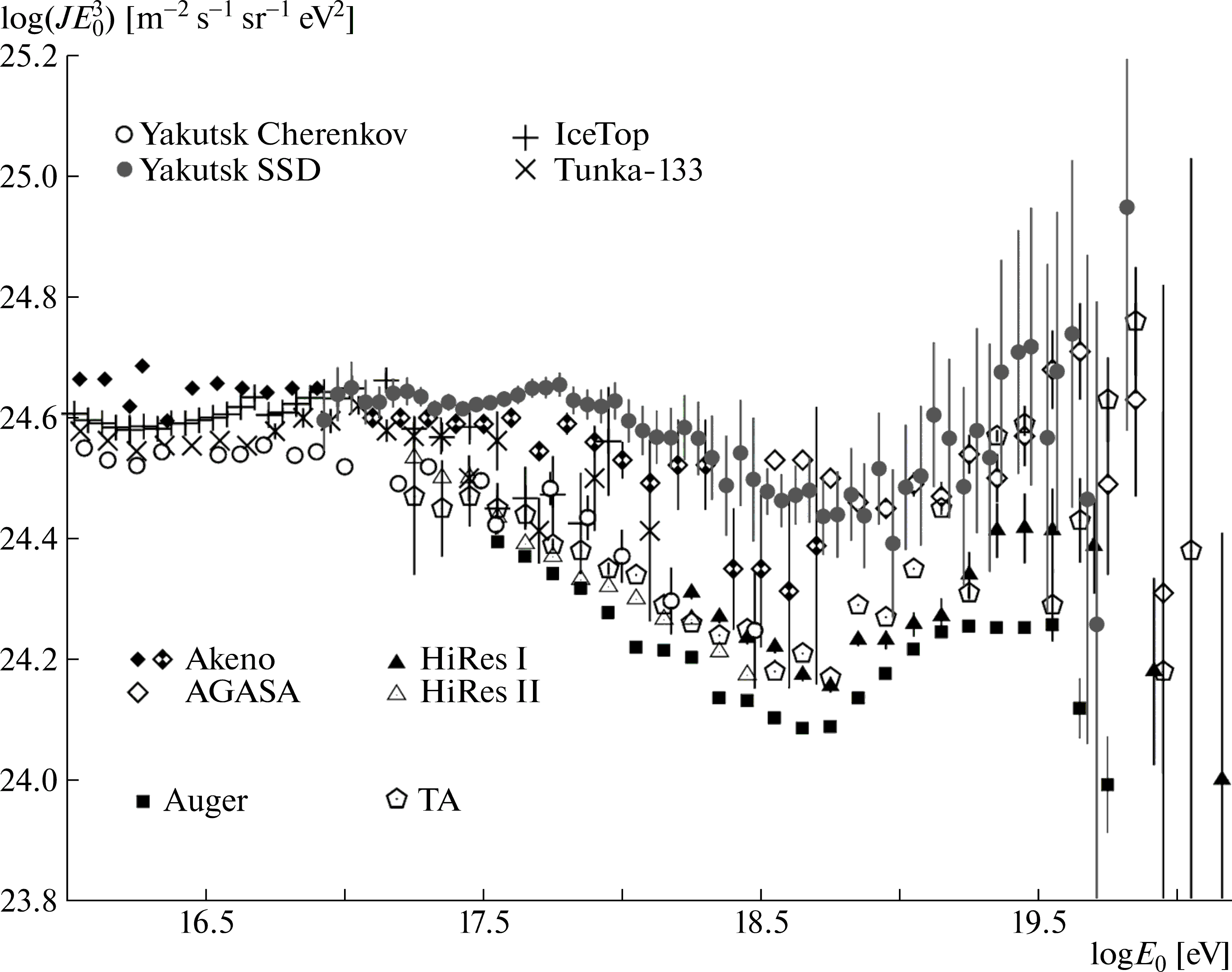}
    \caption{Differential energy spectrum of cosmic rays according to data from various arrays: (closed and open circles) results obtained at the Yakutsk EAS array in the present study and in~\cite{bib:26} from EAS Cherenkov radiation and (closed and half-closed diamonds) Akeno (1984, 1992)~\cite{bib:27, bib:28}, (open diamonds) AGASA~\cite{bib:29}, (inclined and right crosses) Tunka-133~\cite{bib:30} and IceTop~\cite{bib:31}, (closed and open triangles) HiRes-I~\cite{bib:6} and HiRes-II~\cite{bib:33}, (closed boxes) Auger~\cite{bib:7}, and (pentagons) Telescope Array data.}
    \label{fig:6}
\end{figure}

We have considered more than $10^6$ showers detected over the period of continuous operation of the Yakutsk EAS array from 1974 to 2017. The spectrum was constructed on the basis of the procedure proposed in~\cite{bib:25}. The energy of individual events was found according to the refined calorimetric formula (\ref{eq:21}), which depends only slightly on models of EAS development and which relies on results close to one another (see Table~\ref{t:2}). The absorption ranges were taken from the calculations illustrated in Fig.~\ref{fig:3} and performed for the real mixed composition of primary particles~\cite{bib:23, bib:24}. In Fig.~\ref{fig:6}, the resulting spectrum is represented by closed circles. The open circles correspond to the spectrum obtained in~\cite{bib:26} at the Yakutsk EAS array from EAS Cherenkov radiation. The closed, half-closed, and open diamonds stand for Akeno (1984, 1992)~\cite{bib:27, bib:28} and AGASA~\cite{bib:29} data. The inclined and right crosses represent the spectra obtained at, respectively, the Tunka-133~\cite{bib:30} and Ice Top~\cite{bib:31} arrays. The closed and open triangles correspond to HiRes-I~\cite{bib:6} and HiRes-II~\cite{bib:33} data. The closed boxes stand for the Auger (The Pierre Auger Observatory) spectrum~\cite{bib:7}.

Our spectrum agrees with the Akeno–AGASA spectra~\cite{bib:27, bib:28, bib:29} within the experimental errors over the whole range of measured energies. Possibly, this is due to the use of similar scintillation detectors and similar data-analysis procedures in these two cases. Good agreement with Tunka-133~\cite{bib:30} and Ice Top~\cite{bib:31} data is observed at $\E \simeq 10^{17}$~eV. For $\E > 10^{18}$~eV, our results and the results obtained at HiRes~\cite{bib:6, bib:33} and Auger~\cite{bib:7} disagree substantially, possibly because of some special technical features of those arrays.

\section{Conclusions}

The application of the \corsika{} code to the Yakutsk EAS array made it possible to reanalyze critically its energy calibration, which has long been been the subject of lively discussions and disagreement with colleagues performing similar experiments at other arrays worldwide. This became possible owing to the availability of modern models of EAS development. Relying on these models, we were able to calculate responses of scintillation detectors and to obtain, on this basis, a set of possible estimates of the primary energy [see Eqs. (\ref{eq:9})-(\ref{eq:12})]. The calculations revealed that, in expressions (\ref{eq:1}) and (\ref{eq:4}), the energy scattered in the atmosphere in the form of an electromagnetic component is overestimated by 12\% to 17\%, depending on the shower-maximum depth $\xmax$ (Fig.~\ref{fig:4}); in Eq. (\ref{eq:4}), this difference is additionally aggravated by an overestimation of about 17\% shifting the transparency of the atmosphere in the undesirable direction. The new calorimetric result for $\E$ in (\ref{eq:21}) reduced its estimate in relation to that in (\ref{eq:4}) by a factor of about 1.28 and diminished substantially the intensity of the energy spectrum measured at the Yakutsk EAS array (Fig.~\ref{fig:6}).

\begin{acknowledgments}
    This work was supported by the Program of the Presidium of Russian Academy of Sciences High-Energy Physics and Neutrino Astrophysics and by the Russian Foundation for Basis Research (project no. 16-29-13019 ofi-m).
\end{acknowledgments}

\bibliography{refs}

\end{document}